\documentclass[12pt,preprint,dvips]{aastex}

\def\st{\scriptstyle}

\def\be{\begin{equation}}
\def\ee{\end{equation}}
\def\bea{\begin{eqnarray}}
\def\eea{\end{eqnarray}}

\def\approxlt{\ifmmode \rlap{$<$}{}_{{}_{{}_{\textstyle\sim}}} \else%
$\rlap{$<$}{}_{{}_{{}_{\textstyle\sim}}}$\fi}
\def\approxgt{\ifmmode \rlap{$>$}{}_{{}_{{}_{\textstyle\sim}}} \else%
$\rlap{$>$}{}_{{}_{{}_{\textstyle\sim}}}$\fi}
\def\p{\ifmmode\pm\else$\pm$\fi}

\def\msun{\ifmmode \rm{M}_\odot \else M$_\odot$\fi} 
\def\mdot{\ifmmode \dot M \else $\dot M$\fi}    

\def\hide#1{}

\newcounter{sub}
\newcounter{subeqn}[sub]
\renewcommand{\thesubeqn}{\alph{subeqn}}
\renewcommand{\theequation}{\thesub\thesubeqn}

\def\be{\begin{equation}}
\def\ee{\end{equation}}

\def\lp{\left(}
\def\rp{\right)}

\def\ls{\left[}
\def\rs{\right]}

\def\st{\stepcounter{sub}}
\def\stq{\stepcounter{subeqn}}
\def\bea{\begin{eqnarray}}
\def\eea{\end{eqnarray}}

\newcommand\xxi{{\mbox{\boldmath $\xi$}}}
\newcommand\OOmega{\mbox{\boldmath $\Omega$}}

\newcommand\nab{\mbox{\boldmath $\nabla$}}

\def\v{{\bf v}}
\def\r{{\bf r}}

\newcommand\B{{\bf B}}

\newcommand\no{\nonumber}

\begin{document}
\title{X-ray aurora in neutron star magnetospheres}

\author{V. Rezania$^{1,2}$, J. C. Samson, and P. Dobias}

\affil{1- Theoretical Physics Institute,
    Department of Physics,
    University of Alberta\\
    Edmonton, AB, Canada, T6G 2J1}
\affil{2- Institute for Advanced Studies in Basic Sciences,
          Zanjan 45195, Iran}


\begin{abstract}
The discovery of quasi-periodic oscillations (QPOs) in low-mass
X-ray binaries (LMXBs) has been reported and discussed in recent
studies in theoretical and observational astrophysics. The {\it
Rossi X-Ray Timing Explorer} has observed oscillations in the
X-ray flux of about 20 accreting neutron stars. These oscillations
are very strong and remarkably coherent. Their frequencies range
from $\sim 10$ Hz to $\sim 1200$ Hz and correspond to dynamical
time scales at radii of a few tens of kilometers, and are possibly
closely related to the Keplerian orbital frequency of matter at
the inner disk. Almost all sources have also shown twin peaks QPOs
in the kHz part of the X-ray spectrum, and both peaks change
frequencies together.

Bearing in mind these facts, several models explain the observed
frequencies by the oscillation modes of an accretion disk. In the
beat-frequency interpretation, the upper kHz QPOs are associated
with the Keplerian frequency of the accreting gas flowing around a
weakly magnetized neutron star ($B\sim 10^7-10^{10}$ G) at some
preferred orbital radius around the star, while the lower kHz QPOs
result from the beat between the upper kHz QPO and the spin
frequency of the neutron star. In the relativistic precession
model, while the upper kHz QPO is still the orbital frequency of
matter at some radius in the disk, the lower kHz QPOs are
considered as general relativistic precession modes of a free
particle orbit at that radius.

In this study we propose a new generic model for QPOs based on
oscillation modes of neutron star magnetospheres. We argue that
the interaction of the accretion disk with the magnetosphere can
excite resonant shear Alfv\'en waves in a region of enhanced
density gradients. We demonstrate that depending on the distance
of this enhanced density region from the star and the magnetic
field strength, the frequency of the field line resonance can
range from several Hz (weaker field, farther from star), to
approximately kHz frequencies (stronger field, $\sim 6-10$ star
radii from the star).  We show that such oscillations are able to
significantly modulate inflow of matter from the high density
region toward the star surface, and possibly produce the observed
X-ray spectrum. In addition, we show that the observed $2:3$
frequency ratio of QPOs is a natural result of our model.

\end{abstract}
%
%
\section{introduction}

Recent years have seen a surge in interest in studies of accreting
neutron stars and black holes.  One reason is that these objects
provide a very unique window to a deeper understanding of the
physics of strong gravity and dense matter. The complicated
structure of these stellar objects, including strong gravity,
gravitationally-captured plasma accretion, rapid rotation,
bursting in accreted materials, thermal X-ray emission, and
several other features makes a most fascinating object for both
theoretical and observational investigations. One of the most
interesting features of these objects is the interaction of
accreting plasma with the stellar magnetic field. Matter is
transferred from a normal donor star to the compact object. MHD
models of the infalling plasma and the neutron star magnetosphere
allow a relatively simple first approach to an understanding of
plasma flow and the configuration of the magnetosphere around a
neutron star.

Research on plasma accretion by magnetic neutron stars began in
the early 1970s with the discovery of periodic oscillations in
X-ray fluxes of the stars Cen X-3 \citep{Gia71,Sch72} and Her X-1
\citep{Tan72}.   Interest in these systems rose about two decades
ago, after the first observation of quasi-periodic oscillations
(QPOs) in the persistent X-ray fluxes of the luminous X-ray
sources \citep{Van85,Has86,MP86}.  These oscillations are detected
both in low and high frequencies with frequencies ranging from
$\sim 1$ to $\sim 100$ Hz and from $\sim 300 $ to $\sim 1200$ Hz,
respectively.  The {\it Rossi X-Ray Timing Explorer} (RXTE) has
observed persistent X-ray emissions about 20 accreting neutron
stars in low-mass X-ray binaries (LMXBs). Generally, three
different millisecond phenomena have now been observed in X-ray
binaries: the twin kilohertz quasi-periodic oscillations (kHz
QPOs), the burst oscillations, and the spin frequency of the
accreting, low-magnetic field, neutron star.  These kilohertz
frequencies are the highest frequency oscillations ever seen in
any astrophysical object, and are now widely interpreted as due to
orbital motion in the inner accretion flow.  The oscillations
sometimes have one spectral peak.  In most sources, however, the
X-ray spectrum shows two simultaneous kHz peaks with a roughly
constant peak separation, moving up and down together in time as a
function of photon count rate.  In this work, we focus on the
Hz-kHz QPOs.

The burst oscillations with frequencies in the $300-600$ Hz range
have been observed in a number of the X-rays sources, usually
during a type I X-ray burst due to a thermonuclear runaway in the
accreted matter on the neutron star surface.    Furthermore, the
frequency of the burst oscillation increases by $1-2$ Hz during
the burst tail, converging in time to a stable frequency at the
neutron star spin frequency.   It is believed that the burst
oscillations arise due to a hot spot or spots in an atmospheric
layer of the neutron star. This has been explained by a decoupling
of the atmospheric layer from the star and expansion by $5-50$\,m
during the X-ray burst, while its angular momentum remains
constant. Therefore, during the burst tail, the frequency
increases due to spin-up of the atmosphere as it re-contracts. The
asymptotic frequency corresponds to a fully re-contracted
atmosphere and is the closest to the true neutron star spin
frequency (see the recent review by \cite{BS03}).  In a few
sources, however, a rather large frequency drift, $\sim 5-10$ Hz,
is observed. This drift is not consistent with the above model (
see \cite{RS03} for more details).

Although there are some detailed differences in different types of
X-ray sources, the observed QPOs in all sources are remarkably
similar, both in frequency and peak separation.  (Six of the 20
known sources are originally identified as ``Z'' sources and the
rest are known as ``atoll'' sources, see table 1. For more
information about the atoll and Z sources, see \citet{Van00}.)
Such a similarity shows that QPOs should depend on general
characteristics of the X-ray sources which are common to all
systems.  In other words, the QPO can be regarded as a generic
feature of the accreting neutron star.

In the accretion disk of accreting binaries the material from the
donor star moves around the compact star in near-Keplerian orbits.
Depending on the binary separation, the disk radius varies from
$10^5$ to $10^7$ km. Although there is still no certainty about
the flow geometry in the inner emitting region, the extension of a
part of the flow down into the emitting region in the form of a
Keplerian disk is the basic assumption of most proposed models for
accretion of matter onto low magnetic field neutron stars (see for
example \cite{Mil98a}). Due to the interaction with the magnetic
field of the star, radiation drag, and relativistic effects, the
flow will be terminated at a radius $r_{\rm in}$, the innermost
stable orbital radius that is larger than radius of the compact
star $R_s$. Within $r_{in}$ matter may leave the disk and so the
flow is no longer Keplerian. Remarkably, the characteristic
dynamical time scale for materials moving near the compact object
is comparable with the observed millisecond X-ray variability, ie.
$\tau_{\rm dyn}=(r^3/G M)^{1/2}\sim 2~ {\rm ms}~ (r/100 {\rm
km})^{3/2} (1.4 M_\odot/M)^{1/2}$.   Such a natural periodicity is
the foundation of most models for the observed QPOs. We will
discuss these models in the next section.

\begin{figure}[ht]
\centerline{\includegraphics[width=.8\hsize]{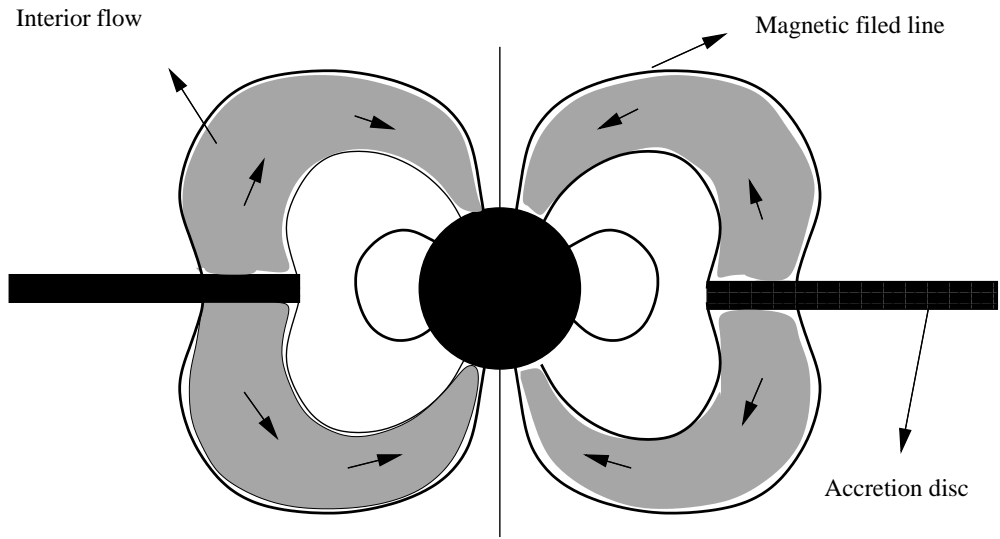}}
\caption{Side view of the magnetic field and accretion disk in
accreting neutron stars.  The neutron star's strong gravity causes
a very high velocity flow toward the magnetosphere.  As a result,
the magnetosphere is pushed inward in the disk plane but balloons
outward in direction away from the disk plane. Some of the plasma
may leave the disk and flow along the field lines.}
\label{star-fig1}
\end{figure}

Another common feature in accreting binaries is the interaction of
the accretion disk with neutron star's magnetosphere.   In all
sources, matters transfer from a normal donor star to the compact
object. This matter is accelerated by the gravitational pull of
the compact object and hits the magnetosphere of the star with a
sonic/supersonic speed. The MHD interaction of the infalling
plasma with the neutron star magnetosphere, will alter not only
the plasma flow toward the surface of the star, as assumed by
current QPO models, but also the structure of the star's
magnetosphere.  The magnetic field of the neutron star is
distorted inward by the infalling plasma of the Keplerian
accretion flow. Since the gravity of the star confines the inward
flow to a small solid angle $\sim 10^{-2}$ ster, the magnetic
field of the star will be more compressed in the disk plane than
in other areas, see Fig. \ref{star-fig1}. Furthermore, in a more
realistic picture, one would expect that the highly accelerated
plasma due to the infall process would be able to penetrate
through the magnetic field lines. Such materials will be trapped
by the magnetic field lines and produce enhanced density regions
within the magnetosphere. However, due to their initial
velocities, they move along the field lines, finally hit the star
surface at the magnetic poles and produce the observed X-ray
fluxes (Fig. \ref{star-fig1}). See \cite{GLP77} for more detail.

Besides modifying the geometry of magnetosphere, the compressional
action of the accretion flow can excite some perturbations in the
enhanced density region.  This can be understood by noting this
fact that the inward motion of the accretion flow will be halted
by the outward magnetic pressure at a certain distance from the
star, the Alfv\'en radius \citep{GLP77}. In other words, the
accretion flow pushes the stellar magnetic field toward the star
until the inward pressure of the infalling plasma $\rho v_r^2/2$
balances with the outward magnetic pressure $B_p^2/8\pi$.  Here
$\rho$ and $v_r$ are the density and radial velocity of the
infalling matter and $B_p$ is the poloidal magnetic field at the
disk plane. Therefore, any instability in the disk at the Alfv\'en
radius would disturb such quasi-equilibrium configuration as well
as the structure of magnetosphere.  As an example, the interaction
of the solar wind with earth magnetosphere excites resonant shear
Alfv\'en waves, or field line resonances (FLRs), along the
magnetic field lines \citep{Sam91}.  As a result, one might expect
that such Alfv\'en waves can be excited by the compressive
accreting plasma in the magnetosphere of an accreting neutron star
with the accretion flow playing the role of the solar wind.

The plan of this paper is as follow: In section \ref{sect:qpo} we
discuss some of the current models for the observed QPOs.
Specifically, we will review the two models, the sonic-point beat
frequency model and the relativistic precession model, in more
detail. We outline an occurrence of the resonant coupling in a
magnetized plasma in section \ref{sect:mag}.  Specifically, we
review the excitation of shear Alfv\'en waves by studying linear
perturbations in MHD.  Next, the occurrence of field line
resonances (FLRs), as a result of a resonant coupling between the
compressional and shear Alfv\'en waves, is discussed.  To
illustrate the basic features of FLRs, the excitation of these
resonances in a rectilinear magnetic field and in a dipolar field
are considered. A dipolar topology is likely appropriate for the
magnetosphere of a neutron star, near the surface.   Depending on
the field line where the resonance occurs the eigenfrequency of
the FLR is in range of several hundred Hz to kHz. In section
\ref{amb}, in order to consider a more realistic model for an
accreting neutron star, we study an influence of the ambient flow
along the field line on the excitation of FLRs. We demonstrate
that in this case the eigenfrequency of the Alfv\'en modes is
modulated by the velocity of the plasma flow. Furthermore, in the
presence of the field aligned flow, the plasma displacement
parallel to the magnetic field lines is non-zero. These
displacements, which are missing in case of zero ambient flow,
might be responsible for modulating the motion of infalling
materials toward the magnetic poles and producing the observed
X-ray fluxes. A possible occurrence of more than one peak in the
power spectrum is also discussed. In addition, we show that the
observed $2:3$ frequency ratio of QPOs is a natural result of our
model. Section \ref{conc} is devoted to summarizing our results
and further discussions.

%
%
\section{The QPO models}\label{sect:qpo}
Although the behavior of observed QPOs in the X-ray spectrum of
accreting neutron stars in LMXBs are somewhat similar, a wide
variety of models have been proposed for QPOs. Most, but not all
of these models, involve orbital motion around the neutron star.
These models include  the beat-frequency models, the relativistic
precession model, and the disk mode models. It is beyond the scope
of the present work to discuss all these models.  We point out,
however, some of the main issues of the most prominent models.
Further information can be found in a review by \cite{Van00}.

%
%

\subsection{The sonic-point beat-frequency model}\label{sect:spbfm}
As discussed earlier, beat-frequency models are based on the
orbital motion of the matter at some preferred radius in the disk.
\cite{AS85} firstly proposed a beat-frequency model to explain the
low-frequency\footnote{Two different Low-frequency ($<100$\,Hz)
QPOs were known in the Z sources, the $6-20$\,Hz so-called normal
and flaring-branch oscillation (NBO; \cite{MP86}) and the
$15-60$\,Hz so-called horizontal branch oscillation (HBO;
\cite{Van85}).} horizontal branch oscillation (HBO) seen in Z
sources (see also \cite{Lam85}).  They used the Alv\'en radius
$r_{\rm A} \sim 1.5 \times 10^6~{\rm cm}~\dot{M}_{17}^{-2/7}
\mu_{26}^{4/7} (M_\odot/M)^{-1/7}$ as the preferred
radius\footnote{Note that at $r_{\rm A}$ we have $\rho v_r^2\sim
B_p^2/4\pi$.}.   Here $\dot{M}_{17}$ is the mass of accretion rate
in units of $10^{17}$ g s$^{-1}$ and $\mu_{26}$ is the stellar
magnetic field in units of $10^{26}$ G cm$^3$.

\cite{Mil96,Mil98a} proposed a beat-frequency model, the so-called
the sonic-point beat-frequency model, based on a new preferred
radius, the sonic radius.  In this model, they assumed that near
the neutron star there is a very narrow region of the disk in
which the radial inflow velocity increases rapidly as radius
decreases. Such a sharp transition in the radial velocity of
plasma flow from subsonic to supersonic happens at the ``sonic
point'' radius, $r_{\rm sonic}$
\footnote{
Recently \cite{Zha04} proposed a model for kHz QPOs based on  MHD
Alfv\'en oscillations.  He introduced a new preferred radius, `quasi
sonic point radius', where the Alfv\'en velocity at this radius matches
the Keplerian velocity. The author suggested that the upper and lower
kHz frequencies are the MHD Alfv\'en wave frequencies corresponding to
different Alfv\'en velocities (or different mass densities).
Although his results have a good agreement
with observations, no mechanism suggested in order to explain the excitation of
Alfv\'en wave oscillations through star's magnetosphere. Further, it is not clear how
these Alfv\'en wave frequencies modulate the X-ray flux coming from the surface of the star.}
.  This radius tends to be near to
the innermost stable circular orbit (ISCO) \footnote{In general
relativity, no stable orbital motion is possible within the
innermost stable circular orbit (ISCO), $R_{\rm ISCO} = 6GM/c^2
\approx 12.5 M_{1.4M_\odot}\,\hbox{km}\,$. The frequency of
orbital motion at the ISCO, the highest possible stable orbital
frequency, is $\nu_{\rm ISCO} \approx (1580/M_{1.4M_\odot})\,
\hbox{Hz}$.}, $r_{\rm ISCO}$, however, radiative stresses may
change its location, as required by the observation that the kHz
QPO frequencies vary.  Comparing the HBO and kHz QPO frequencies,
clearly $r_{\rm sonic}\ll r_{\rm A}$, so part of the accreting
matter must remain in near-Keplerian orbits well within $r_{\rm
A}$.

In order to explain how such orbital frequencies can modulate the
X-ray fluxes, they assumed that in the accreting gas some local
density inhomogeneities ``clumps'' are created by
magneto-turbulence, differential cooling of matters, and radiation
drags.   At $r_{\rm sonic}$ these orbiting clumps gradually
accrete onto the neutron star, following a fixed spiral-shaped
trajectory in the frame corotating with their orbital motion and
hit the star surface before they are completely sheared or
dissipated by radiation drag. At the ``footpoint'' of a clump's
spiral flow the matter hits the surface of the star and so the
emission is enhanced. The clump's footpoint travels around the
neutron star at the clump's orbital angular velocity, so a distant
observer whose line of sight is inclined with respect to the disk
axis sees a hot spot that is periodically occulted by the star
with the Keplerian frequency at $r_{\rm sonic}$. This produces the
upper kHz peak at $\nu_2$. The high coherency of the QPO implies
that all clumps are near one precise radius and live for several
$0.01$ to $0.1$ s, and allows for relatively little fluctuations
in the spiral flow. To explain twin peaks, it is also assumed that
some of the accreting plasma is channelled by the magnetic field
onto the magnetic poles that rotate with the star. The beat
frequency at $\nu_1$ occurs because a beam of X-rays generated by
accretion onto the magnetic poles sweeps around at the neutron
star spin frequency $\nu_s$ and hence irradiates the clumps at
$r_{\rm sonic}$ once per beat period, which modulates, at the beat
frequency, the rate at which the clumps provide matter to their
spiral flows and consequently the emission from the footpoints.

An important prediction of the sonic-point model is that
$\Delta\nu = \nu_2-\nu_1$ be constant at $\nu_s$, which is
contrary to observations \citep{LM99}.  However, one obtains
better consistency with observations by assuming that the orbits
of clumps are gradually spiraling down.  As a result, the observed
beat frequency will be higher than the actual beat frequency at
which beam and clumps interact.  This can be understood by noting
that during the clump's lifetime the inflow time of matter from
clump to surface gradually diminishes.  Therefore, the lower kHz
peak gets closer to the upper one, and so $\Delta\nu$ decreases
\citep{LM99}. Such a decrease is enhanced at higher X-ray
luminosity $L_x$ because of a stronger radiation drag causes a
faster spiraling-down, as observed \citep{LM99}.   Due to
complications, it is hard to predict how exactly this affects the
relation between frequencies which makes testing the model more
difficult.

%
%

\subsection{The relativistic precession model}\label{sect:rpm}

Based on general relativity, it is well known that the motion of a
free-particle in an inclined eccentric orbit around a spinning
object experiences both relativistic periastron precession similar
to Mercury's \citep{Ein15}, and nodal precession (a wobble of the
orbital plane) due to relativistic frame dragging \citep{TL18}.
Noting these facts, \cite{SV98,SV99} proposed the relativistic
precession model in which the high kHz QPO frequency $\nu_2$ is
identified with the Keplerain frequency of an orbit in the disk
(similar to the beat-frequency model) and the low kHz QPO
frequency $\nu_1$ with the periastron precession of that orbit.
They also compared the nodal precession of the same orbit to the
frequency of one of the observed low-frequency ($\nu_{LF}\sim
10-100$\,Hz) in LMXBs.

In this model the kHz peak separation varies as
$\Delta\nu=\nu_2(1-6GM/rc^2)^{1/2}$ caused by the relativistic
nodal precession
and the relativistic periastron precession \citep{SV98,SV99,ML98}.
Here $I$ is the star's moment of inertia and $r$ the orbital radius.
\cite{MS99} and \cite{Ste99} noticed that stellar oblateness affects both
precession rates and must be corrected for.   Using acceptable neutron star
parameters, they obtained an approximate match
with the observed $\nu_1$, $\nu_2$ and $\nu_{LF}$ relations provided $\nu_{LF}$
is {\it twice} (or perhaps sometimes four times)\footnote{It is
suggested that such a relation
could in principle arise from a warped disk geometry \citep{MS99}.}
the nodal
precession frequency
\citep{MS99}.  However, to get a more precise match between model
and observations,
one needs to use additional free parameters \citep{SV99}.

In this model $\Delta\nu$ and $\nu_{s}$ are not expected to be
equal as in a beat-frequency interpretations. An interesting
result of the relativistic precession model is that $\Delta\nu$
should decrease not only when $\nu_2$ increases (as observed) but
also when it sufficiently decreases.

Although the relativistic precession model explains the three most
prominent frequencies observed in the X-ray spectrum of LMXBs,
with the three main general-relativistic frequencies
characterizing a free-particle orbit, there are still some
unanswered questions. For example, how stable are precessing and
eccentric orbits in a disk, how is the flux modulated at the
predicted frequencies, and why does the orbital frequency $\nu_2$
change with X-ray luminosity?. Furthermore, from this model
relatively high neutron star masses ($1.8-2$\msun), relatively
stiff equations of state, and neutron star spin frequencies in the
$300-900$\,Hz range are obtained.


\section{The magnetospheric model}\label{sect:mag}
\subsection{Field line resonances in the earth's magnetosphere}

In the last few decades there was a large expansion of the
exploration of the near-Earth, plasma environment. One of the
recent advances in the space plasma physics is the proof of the
connection between FLRs and auroral dynamics \citep{Sam96}.
\cite{Sam03} use a combination of ground based and satellite
observations with analytical and computational models to
demonstrate a connection between FLRs and the formation of auroral
arcs.  \cite{Dob04} outline a possibility that the FLRs are a
triggering mechanism for the nonlinear plasma instabilities that
occur in the near-Earth magnetotail during magnetospheric
substorms. Since FLRs are a generic feature exciting in magnetized
plasmas with gradients in the Alfv\'en velocity and with
reflection boundaries, they are likely to occur in accreting
neutron star magnetospheres as well. In the following sections we
outline the occurrence of the resonant coupling within the ideal
MHD approach and illustrate the resonance mechanism with some
simple examples. We present results of computational models of
shear Alfv\'en waves in dipolar magnetic field, and we discuss the
effect of an ambient flow parallel to the field.


\subsection{Magneto-hydrodynamic waves}

In this section we study MHD and applications to Alfv\'en waves
and FLRs. In general, the dynamics of a magnetized plasma is
described by plasma density $\rho=m n$, plasma pressure $p$,
gravitational potential $\Phi$, velocity vector $\v$ and magnetic
field $\B$: \st\begin{eqnarray}\label{eqm} \stq\label{euler}
&&\partial\v/\partial t + \v \cdot \nab \v =
 - \nab p/\rho  + (1/4\pi\rho)\B \times (\nab \times \B) - \nab\Phi,\\
\stq\label{continuty}
&&\partial\rho/\partial t + \nab\cdot(\rho \v)=0,
\end{eqnarray}
An adiabatic equation of state with an adiabatic index $\gamma$,
$p/\rho^\gamma=$const., is assumed in this paper in order to
complete the above equations.

In a steady state (ie. $\partial/\partial t=0$), Eqs. (\ref{eqm})
have been studied in detail regarding with the problem of stellar
winds from rotating magnetic stars \citep{Mes61,Mes68} and in
connection with diffusing/flowing plasma into magnetospheres in
accreting neutron stars \citep{EL77,EL84,GLP77}. Obviously, in
case of accreting neutron stars, $\v$ represents the inflow
velocity of matter accreted to stars, while in stars with a
stellar wind it represents the plasma outflow velocity. By
decomposing the velocity and magnetic field vectors into poloidal
and toroidal components: \st \be\label{comp} \v=\v_p + \Omega
\varpi \hat{\phi},~~~~~~~\B=\B_p + B_\phi\hat{\phi}, \ee one can
obtain \bea \st\label{vp}
&&\v_p=(f/\rho)\B_p,\\
\st\label{om} &&\Omega=\Omega_s+(f/\rho)(B_\phi/\varpi), \eea
where $f$ is the mass flux along a magnetic flux tube of unit flux
and $\Omega_s=\OOmega_s\cdot \hat{z}$ is the angular velocity of
the star \citep{GLP77}. The subscripts $p$ and $\phi$ denote
poloidal and toroidal components, respectively, $\Omega$ is the
angular velocity of plasma at a distance $\varpi$ from the axis of
the aligned rotator, and $\hat{\phi}$ is a unit toroidal vector.
We note that both $f$ and $\Omega$ are constant along a given
field line. Equation (\ref{om}) can be rewritten as \st \be
\Omega=\Omega_s+(v_p/\varpi)(B_\phi/B_p), \ee where $v_p=\v_p\cdot
\hat{p}$ is the magnitude of velocity along the poloidal magnetic
field $\B_p$ with magnitude of $B_p$. Furthermore, \cite{GLP77}
have shown that at distances close to the Alfv\'en radius, the
poloidal component of the inflow velocity (for accreting systems)
$v_p$ approaches the Alfv\'en velocity $v_{\rm A}$, ie. \footnote{
In order to estimate the poloidal component of the inflow velocity
$v_p$, one needs to integrate the momentum Eq. (\ref{euler}): \st
\be\label{vp_1} (1/2)(v_p^2 + \Omega^2\varpi^2)-\Omega_s\Omega
\varpi^2 -GM/r={\rm const}, \ee where $M$ is the mass of the
neutron star \citep{Mes68,GLP77}. In above equation the pressure
term is neglected.  Equation (\ref{vp_1}) shows conservation of
energy in a corotating frame with the star, while in a nonrotating
frame the extra term $\Omega_s\Omega\varpi^2$ appears, that
represents the work done by the magnetic field on the flowing
plasma. However, as argued by \cite{GLP77} the magnitude of $v_p$
inside of the magnetosphere is nearly equal to the free-fall
velocity, ie. \st \be\label{vp1} v_p\sim (2GM/r)^{1/2}. \ee } \st
\be v_p^2(r_{\rm A})=v_{\rm A}^2(r_{\rm A}). \ee

The possible perturbations of a magnetized plasma and MHD waves
are found by specifying the equilibrium configuration of the star
and then solving Eqs. (\ref{eqm}). This is a nontrivial problem,
that is addressed by several investigators. However, in this paper
we will be interested in propagation of shear Alfv\'en waves in
the star's magnetosphere and their resulting FLRs.


\subsection{Field line resonances}

In MHD, waves can propagate in three different modes including
shear Alfv\'en waves, and the fast and slow compressional modes
\citep{LL92}. In a homogeneous plasma, one can easily show that
these three modes are linearly independent. However, in an
inhomogeneous media these three modes can be coupled, yielding
either a resonant coupling \citep{Sou74,Has76}, or an unstable
ballooning mode \citep{OT93,Liu97}. FLRs result from the coupling
of the fast and the shear Alfv\'en modes.

The linear theory of the FLRs was developed by \cite{CH74} and
\cite{Sou74}, and applied to auroral phenomena by \cite{Has76}.
\cite{Sam03} developed a nonlinear model with a nonlocal
conductivity model to explain the evolution of field aligned
potential drops and electrical acceleration to form  aurora. They
studied the possible coupling between the fast compressional mode
and the shear Alfv\'en mode in an inhomogeneous plasma with radial
gradient in the Alfv\'en velocity $v_{\rm A}=B/\sqrt{4\pi \rho}$.
For further discussion of the field line resonances see for
instance \cite{Sti92}, and for the example of the numerical
simulation see \cite{RW94}.

To illustrate the resonant coupling between the fast compressional
mode and the shear Alfv\'en mode we study the excitation of FLRs
in an inhomogeneous plasma in two simple models for the magnetic
field configuration, a rectilinear magnetic field model and a
dipolar magnetic field model.


\subsubsection{FLRs in a rectilinear magnetic field}
Plasma dynamics can often be described by the ideal MHD equations
as: \st
\begin{eqnarray}\label{mhd_eq}
\stq
&&\rho( \partial/\partial t +\v\cdot\nab)\v =
- \nab p + \B \times (\nab\times \B), \label{mhd_mom} \\
\stq
&&\partial \B/\partial t = \nab\times (\v \times \B), \label{mhd_ind} \\
\stq
&&\partial \rho/\partial t + \nab \cdot (\rho \v) = 0, \label{mhd_cont} \\
\stq
&&d(p\rho^{-\gamma})/dt=0. \label{mhd-stat}
\end{eqnarray}
where we neglect the effect of gravitational attraction on the
plasma here, see Eqs. (\ref{eqm}) for more detail. Introducing
Eulerian perturbations to the ambient quantities by
\begin{eqnarray}
&& p = p_0 + \delta p,~~ \rho = \rho_0 +\delta \rho, \no\\
&&\B = \B_0 + \delta \B,~~ \delta\v = \partial\xxi/\partial t,\no
\end{eqnarray}
one can derive the linear wave equation by simplifying Eqs.
(\ref{mhd_eq}) up to the first order in the perturbations. We
choose coordinates such that the ambient magnetic field is in the
z-direction.  Without loss of generality, we assume that the
gradients in the plasma parameters are in the x-direction. Setting
the plasma displacement as \st
\begin{equation}
\xxi (\r,t) = \xxi(x) e^{-i (\omega t - k_y y - k_z z)},
\end{equation}
Eqs. (\ref{mhd_eq}) reduce to \citep{HS92}
\st\begin{equation} \label{har}
\frac{d^2 \xi_x}{dx^2} + \frac{F'(x)}{F(x)} \frac{d \xi_x}{dx} + G(x) \xi_x = 0,
\end{equation}
with
\st\begin{equation}\label{G}
G(x) = \frac{\omega^2 \Bigg(\omega^2 - (v_{\rm S}^2+v_{\rm A}^2)(k_y^2 + k_z^2)\Bigg) +
k_z^2 (k_y^2 + k_z^2) v_{\rm A}^2 v_{\rm S}^2}{(v_{\rm S}^2+v_{\rm A}^2)\omega^2 -
k_z^2 v_{\rm A}^2 v_{\rm S}^2},
\end{equation}
and
\st\begin{equation}\label{F}
F(x) = \frac{\omega^2 - k_z^2 v_{\rm A}^2}{G(x)},
\end{equation}
where $v_{\rm A}^2(x) = B^2/4\pi\rho$ and $v_{\rm S}^2(x) = \gamma
p/\rho$, and $'=d/dx$.  Equation (\ref{har}) yields two turning
points at $G(x) = 0$ (the compressional wave turns from
propagating to evanescent), and two resonances at $F(x) = 0$.
Close to the resonance positions, Eq. (\ref{har}) can be
approximated by \st
\begin{equation}
\frac{d^2 \xi_x}{dx^2} + \frac{1}{x - x_0} \frac{d \xi_x}{dx} + G(x) \xi_x = 0,
\end{equation}
where $x_0$ is the position that the resonance occurs.

In case of a strong magnetic field, or for a cold plasma (ie.
$p_{\rm fluid}/p_{\rm magnetic}\ll 1$), one can put $v_{\rm S}
\simeq 0$. Then $G(x)$ and $F(x)$, (\ref{G}) and (\ref{F}), reduce
to \st
\begin{equation}
G(x) = \frac{\omega^2}{v_{\rm A}^2} - k_y^2 - k_z^2
\end{equation}
and
\st
\begin{equation}\label{F1}
F(x) = \frac{v_{\rm A}^2 (\omega^2 - k_z^2 v_{\rm A}^2)}{\omega^2 -
(k_y^2 + k_z^2) v_{\rm A}^2}.
\end{equation}
In this case Eq. (\ref{har}) has only one resonance (Alfv\'en resonance) at
\st
\begin{equation}\label{flres}
\omega^2 - k_z^2 v_{\rm A}^2=0,
\end{equation}
which corresponds to the dispersion relation for the shear
Alfv\'en wave along the field line.  At the resonance the incoming
compressional wave  reaches field line (or resonant magnetic
shell) upon which the eigenfrequency of the standing shear
Alfv\'en wave (boundaries at the surface of the neutron star) is
the same as the frequency of the incoming compressional, fast mode
fluctuation.

The FLR mechanism is generic and likely to occur in many
astrophysical magnetosphere. As a result, one would expect that
the FLRs likely occure not only in the Earth's magnetosphere but
also in the magnetospheres of accreting neutron stars. In the case
of the Earth's magnetosphere, the source of energy for the
resonant interaction is the interaction of the solar wind with the
magnetosphere \citep{HS92}. In accreting neutron stars, the
accreted plasma interacts with the stars' magnetosphere, allowing
the compressional mode to propagate into the magnetosphere and
flow along the field lines toward the magnetic poles. Such a
compressional action of the accretion flow can excite resonant
shear Alfv\'en waves in the enhanced density regions filled by
plasma flowing along the field lines. In section \ref{amb} we
consider this mechanism more carefully to address its possible
relation to those quasi periodic oscillations observed in
accreting neutron stars in LMXBs.


\subsubsection{FLRs in a dipolar field} \label{numdip}

In previous section we studied the excitation of FLRs in the
simple box model (the rectilinear magnetic field). In this
section, however, we consider a more realistic configuration, the
dipolar magnetic field.  Analytic solutions for FLRs in dipole
fields can be found in \cite{TW84,Sam96}.  \cite{RW94} outline
computational (numerical) methods for modeling FLRs.

\begin{figure}[h]
\begin{center}
\includegraphics[angle=0,width=10cm]{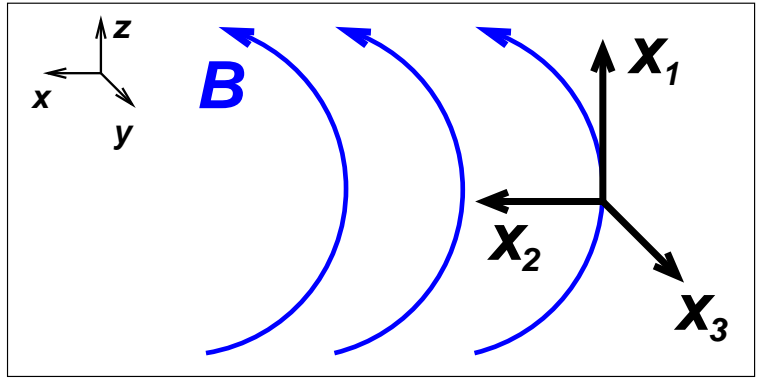}
\caption{Sketch of curvilinear coordinates for curvilinear or
dipole magnetic field configurations.} \label{cl-coor}
\end{center}
\end{figure}

Let coordinates ($x_1$, $x_2$, $x_3$), such that the coordinate
$x_{1}$ is directed along field lines, the coordinate $x_{2}$ is
in the direction of the radius of curvature, and the coordinate
$x_{3}$ is in the azimuthal direction, see Fig. \ref{cl-coor}
\footnote{In dipolar coordinates, or coordinates close to dipolar,
a notation ($\mu, \nu, \phi$) is
usually used where $\mu=\cos\theta/r^2$, $\nu=\sin^2\theta/r$, and ($r$, $\theta$, $\phi$)
are spherical
coordinates .}. The equation of motion can be written in
components as
\begin{eqnarray}
\rho \partial_{tt} \xi_{1} &=& 0, \label{on1} \\
\rho \partial_{tt} \xi_{2} &=& \frac{B}{h_{1}h_{2}} \ls \partial_{2}
(\partial_{2} B h_{3} \xi_{2} + \partial_{3} B h_{2} \xi_{3}) + \partial_{1}
\lp \frac{h_{2}}{h_{1}h_{3}} \partial_{1} B h_{3} \xi_{2}\rp\rs, \label{on2} \\
\rho \partial_{tt} \xi_{3} &=& \frac{B}{h_{1}h_{3}} \ls \partial_{3}
(\partial_{2} B h_{3} \xi_{2} + \partial_{3} B h_{2} \xi_{3}) + \partial_{1}
\lp \frac{h_{3}}{h_{1}h_{2}} \partial_{1} B h_{2} \xi_{3}\rp \rs, \label{on3}
\end{eqnarray}
where $h_1$, $h_2$, $h_3$ are the metric (Lam\'e) coefficients. We
limited ourselves to a 2-dimensional problem in the $x_1x_2$ (or
$xz$) plane.  Here $\partial_t=\partial/\partial t$ and
$\partial_i=\partial/\partial x_i$.

\begin{figure}[h]
\begin{center}
\includegraphics[angle=0,width=12cm]{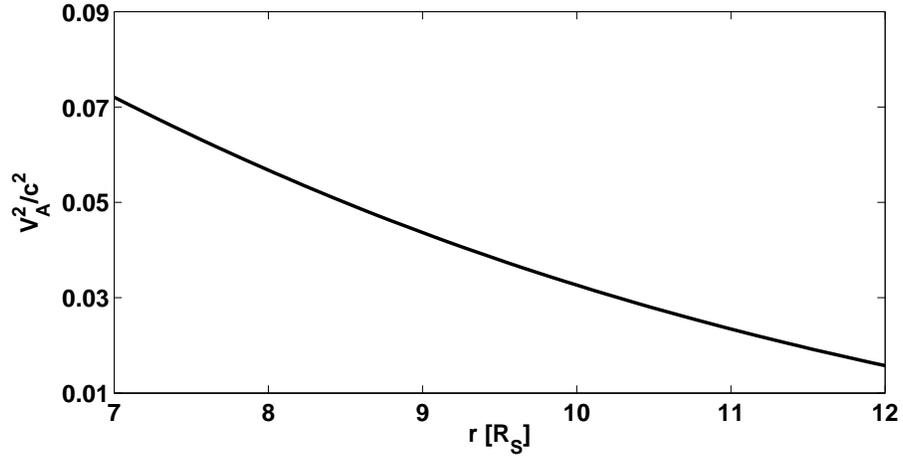}
\caption{$v_{\rm A}^2$ in equatorial plane} \label{dip_va2e}
\end{center}
\end{figure}

\begin{figure}[h]
\begin{center}
\includegraphics[angle=0,width=12cm]{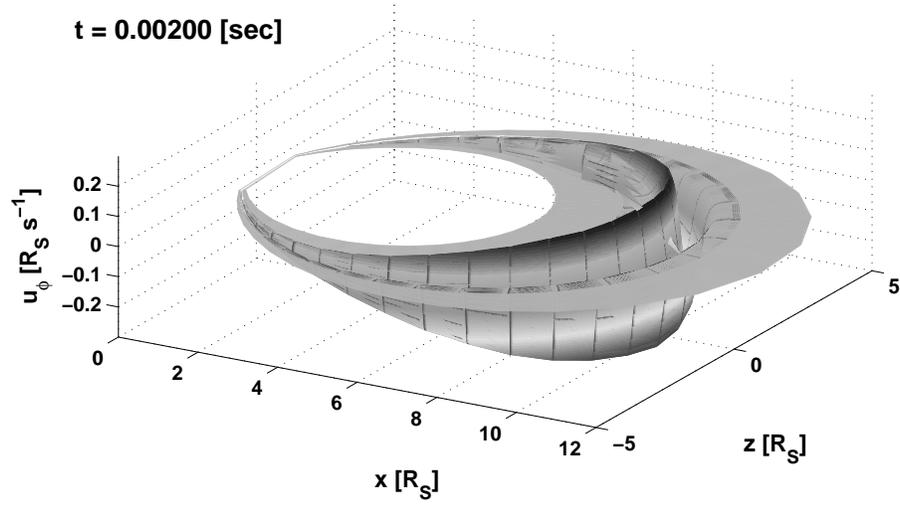}
\caption{Azimuthal velocity, $v_\phi$, profile for a FLR in a
dipole, neutron star's magnetosphere.  $R_s$ is in units of the
star radius.} \label{dip_ini}
\end{center}
\end{figure}

In a strong magnetic field the usual definition of the Alfv\'en
velocity, $v_{\rm A}^2=B^2/4\pi\rho$, is failed by resulting
values larger than speed of light, $c$. To avoid of such confusion
we define the relativistic Alfv\'en velocity by using enthalpy \st
\begin{equation}\label{enthalpy}
h = \frac{c^2 \rho + B^2/4\pi}{c^2}
\end{equation}
rather than the plasma density $\rho$. So the Alfv\'en velocity
shall be $v_{\rm A}^2 = B^2/4\pi h$ (Fig. \ref{dip_va2e}). As
shown in Fig. \ref{dip_va2e} close to the star the relativistic
Alfv\'en velocity saturates at the speed of light, ie. $v_{\rm A}
\sim c$.  Such definition for the Alfv\'en velocity is extensively
used in studying of the relativistic jets dynamics in quasars.
Figure \ref{dip_ini} shows a computational model of the profile of
the azimuthal component of the velocity. In a FLR with small
azimuthal wavenumber, the displacement and velocity fields are
predominantly azimuthal, with a $180^o$ phase shift across the
position of the resonance.

%
%

\section{Magnetohydrodynamics in the presence of ambient flow}\label{amb}

As discussed above, FLRs have been used to model auroral
observations in the Earth's magnetosphere. Although the occurrence
of these resonances is generic and likely to be excited in any
magnetosphere with input of compressional of energy, one must
carefully evaluate the differences between the Earth and neutron
star magnetospheres.  In the case of the Earth's magnetosphere,
due to the small gravitational attraction of the Earth and also
its large distance from the Sun, the solar wind more or less hits
the whole Sunward side of the geomagnetosphere, and produces the
so-called bow shock structure at the outer boundary. However, the
strong gravity of the neutron star creates a supersonic converging
flow long before the flow hits the star's magnetosphere. Such a
localized flow is able to change the structure of the
magnetosphere in local areas particularly in the equatorial plane.
In addition, the highly variable nature of the exterior flow can
change the magnetosphere's structure with time dramatically.  The
large flux of plasma stresses the outer star's magnetosphere and
creates a relatively high plasma density throughout the magnetic
field.  The plasma then flows along the field lines, an interior
flow, until it hits the star's surface near the magnetic poles,
see Fig. \ref{star-fig1}. Furthermore, the strong magnetic field
of the star, the rapid rotation, and the intense radiation
pressure from the surface must be added to the problem.

One of the most prominent differences between magnetospheres of
accreting neutron stars and the Earth's magnetosphere is the
presence of an ambient flow along the field lines where the
resonance takes place \citep{GL91}.   In this section we study the
excitation of FLRs by considering such plasma flow in the
magnetosphere.

The presence of a flow $\v$ in the plasma adds more modes to the
plasma waves. In general, such flow is a combination of plasma
flow along the magnetic field lines, $\v_p$, and rotational motion
of the plasma around the star with angular velocity $\OOmega$, ie.
$\v=\v_p+\OOmega\times\r$, see Eq. (\ref{comp}). The existence of
$\v_p$ itself allows a non-zero parallel component (relative to
the magnetic field) of displacement that vanishes in a cold plasma
with $\v_p=0$. We note that this component is responsible for
modulating the infalling plasma flow toward the star surface with
FLRs' frequencies and producing the observed X-ray fluxes, see Eq.
(\ref{par-comp}) below. We will back to this point later.

As we mentioned earlier, due to the relatively strong magnetic
field in accreting neutron stars one needs to consider the
relativistic Alfv\'en velocity as \st
\begin{equation}\label{rev_alfv}
v_{\rm A}^2 = \frac{c^2 }{c^2 \rho + B^2/4\pi}~\frac{B^2}{4\pi}\,.
\end{equation}
See Eq. (\ref{enthalpy}) for the definition of the enthalpy. Since
the plasma pressure in neutron stars is negligible compared to
magnetic pressure \citep{GLP77} we are using a cold plasma
approximation in this section. Again we use a linear approximation
to obtain a dispersion relation with resonant coupling. Since the
gravity acts in a direction nearly perpendicular to field line for
most of the infall of the matter, we can assume that the flow is
nearly constant. This simplifies the derivation significantly.

The linearized magnetohydrodynamic
equations in the presence of an ambient flux can be obtained from Eqs. (\ref{mhd_eq}) as
\st\begin{eqnarray}\label{eq-mhd1}
\stq\label{xi}
&&h \lp {\partial \delta \v \over  \partial t}+ {\partial \v \over \partial t} + \v \cdot \nab
\delta\v +  \delta\v \cdot \nab \v\rp =
 - \nab \delta p   + {\nab p\over h}\delta h \no\\
&&\hspace{5cm}+ {1\over 4\pi}\delta \B \times \nab \times \B + {1\over 4 \pi}\B
\times \nabla \times \delta \B, \\
\stq
&&{\partial \delta \B\over\partial t} = \nab \times (\delta \v \times \B + \v
\times \delta \B), \\
\stq\label{eqs-2}
&&({\partial \over \partial t} + \v\cdot\nab)\delta p + \delta \v \cdot \nab p = - \gamma
( \delta p \nab \cdot \v + p  \nab \cdot \delta \v),
\end{eqnarray}
where $\delta \v=\partial \xxi/\partial t$. The time scale of the
ambient flow variations is much longer than time scale of the
excited MHD perturbations, so $\partial\v/\partial t =0$.
Furthermore, we consider the slow rotation approximation and so
neglect the  toroidal field $\B_\phi$, Eq. (\ref{comp}), to avoid
complexities.   Such assumptions may not meet the actual
configuration precisely.  Nevertheless, these assumptions simplify
our calculations significantly.

Separating Eqs. (\ref{eq-mhd1}) into parallel and perpendicular
components relative to the ambient magnetic field and assuming
perturbed quantities in the form of \st
\begin{equation}
\delta (\r,t) = \delta (\r_\perp) e^{-i (\omega t - k_{||} x_{||})}\,,
\end{equation}
we find
\st \bea\label{eq-mhd2}
\stq\label{par-comp}
&&(-i\omega+ik_{||}v_{||})\xi_{||} =
-\frac{k_{||}v^2_{\rm s}}{\omega-k_{||}v_{||}}\nab\cdot\xxi -
(\xxi_\perp\cdot\nab_\perp)v_{||} \,,\\
\stq\label{perp-comp} &&((\omega-k_{||}v_{||})^2-k_{||}^2v_{\rm
A}^2) \xxi_\perp  =
\frac{\omega-k_{||}v_{||}}{\omega}\frac{\nab_\perp\delta P}{h} -
\frac{\omega-k_{||}v_{||}}{\omega}\frac{\nab_\perp p}{h^2}\delta h
\,,\\
\stq
&&(-i \omega + ik_{||}v_{||})\delta B_{||}= i\omega
B(\nab_\perp\cdot \xxi_\perp) + i\omega
(\xxi_\perp\cdot\nab_\perp) B
+(\delta\B_\perp\cdot\nab_\perp) v_{||}\,,\\
\stq
&&(-i \omega + ik_{||}v_{||})\delta\B_\perp = \omega k_{||} B \xxi_\perp\,, \\
\stq
&&(-i \omega + ik_{||}v_{||})\delta p=i\omega (\xxi_\perp\cdot\nab_\perp)p
+\gamma (-\omega k_{||}p\xi_{||} +i\omega p
\nab_\perp\cdot\xxi_\perp)\,,\\
\stq\label{PP} &&-\frac{\delta
P}{h}=\frac{\omega}{\omega-k_{||}v_{||}}(v_{\rm s}^2+v_{\rm
A}^2)\nab\cdot\xxi -\frac{i\omega k_{||} v_{\rm
A}^2}{\omega-k_{||}v_{||}}\xi_{||} +\frac{\omega k_{||} v_{\rm
A}^2}{(\omega - k_{||} v_{||})^2}
(\xxi_\perp\cdot\nab_\perp)v_{||}\,, \eea
where $\omega$ is the
eigenfrequency, $k_{||}$ is the wave number in the parallel
direction, $\v_p=v_{||}\B/B$, $v_{\rm s}=\sqrt{\gamma p/h}$ is the
sound velocity, and $P=p+B^2/8\pi$ is the total plasma pressure
(fluid $+$ field)\footnote{In order to obtain Eq. (\ref{PP}) we
assumed that at equilibrium $\nab(p+B^2/8\pi)\approx 0$. This is
valid for a very slowly rotating star.}. Note that we assumed
that all ambient quantities are function of $\r_\perp$ only.
Equation (\ref{par-comp}) shows that the plasma displacement
parallel to the ambient field does not vanish even in the cold
plasma approximation ($p=0$)\footnote{In case of zero ambient
flow, $v_{||}=0$, the parallel displacement $\xi_{||}$ vanishes in
the cold plasma}. The non-zero $\xi_{||}$ can affect and then
modulate the motion of plasma along the field lines. Such
modulation will occur at frequency of the resonant shear Alfv\'en
waves $\omega$ as seen in the observed X-ray fluxes.

To analyze the problem analytically, we consider here again the
rectilinear magnetic field configuration.  We note that although
this configuration may not be suitable for the accreting neutron
star, it provides us a descriptive picture that can be applicable
to QPOs.   Setting the magnetic field in z-direction with
gradients in the ambient parameters in the x-direction,
$v_{||}(\r_\perp)=v_p(x)$, and \st
\begin{equation}
\delta (\r,t) = \delta(x) e^{-i (\omega t - k_y y - k_z z)},
\end{equation}
Eqs. (\ref{eq-mhd2}) in the cold plasma approximation ($p=0$) can
be combined into one differential equation in the form of Eq.
(\ref{har})\footnote{ Note that in order to simplify our
calculations we neglect the last term in the RHS of Eq.
(\ref{PP}).} \st\begin{equation} \label{line} \frac{d^2
\xi_x}{dx^2} + \frac{\Xi '(x)}{\Xi (x)} \frac{d \xi_x}{dx} +
\frac{(\omega-k_z v_p)^2 - (k_y^2+k_z^2)v_{\rm A}^2}{v_{\rm A}^2}
\xi_x = 0,
\end{equation}
where \st\be\label{Xi} \Xi(x) = \frac{v_{\rm A}^2}{\omega -
k_z v_p}~ \frac{(\omega - k_z v_p )^2 - k_z^2 v_{\rm A}^2}
{(\omega - k_z v_p )^2 - (k_y^2 + k_z^2)  v_{\rm A}^2 }\,. \ee
Note that here $v_p=v_p(x)$ and $v_{\rm A}=v_{\rm A}(x)$ are
function of $x$.  For $v_p=0$ Eq. (\ref{Xi}) reduces to one
obtained in section 3.3, Eq. (\ref{F}), as expected. The condition
$\Xi (x) = 0$ yields the following modes:
\be
\st\label{om_A}
\omega^{\pm}=  k_z (v_p \pm v_{\rm A}) .
\ee
The resulting modes are combination of a
compressional mode $k_zv_p$ due to the plasma flow and a shear
Alfv\'en mode $k_zv_{\rm A}$.   They produced two frequency peaks
in the spectrum that vary with time if the velocities change with
time. The value of $k_z \sim \pi/L$ where $L$ is the length of
field line and is in order of radius of neutron star, $L \sim
R_s$. Therefore, for $R_s\sim 10^6$ cm and $v_{\rm A}\sim v_p\sim
3\times 10^8$ cm s$^{-1}$ we get $\omega^+ \sim 1000$ Hz that is
comparable with frequencies of the observed quasi-periodic
oscillations. In the limit $v_p\ll v_{\rm A}$ these modes become a
single mode of a regular Alfv\'en resonance. In the case of
$v_{\rm A}\ll c$ these modes become consistent with a single mode
(\ref{flres}).  For a superalfv\'enic flow, ie. $ v_{\rm A}\ll
v_p$, however, the MHD Alfv\'en waves and the resulting FLRs are
suppressed due to propagation of the hydrodynamical wave,
$k_zv_p$.  Therefore, no FLRs are likely to occur at this level.
This can be understood from Eq. (\ref{Xi}) that for
superalfv\'enic plasma flow reduces to
\st\be
\Xi(x) \simeq
\frac{ v_{\rm A}^2}{\omega - k_z
v_p}\frac{k_z^2}{k_y^2+k_z^2}\,, \ee
where as $\omega\rightarrow
k_z v_p$, $\Xi(x)\rightarrow \infty $. As a result, we expect that
the FLR occurs where $v_p$ and $v_{\rm A}$ are comparable. Using
the flow velocity and the shear Alf\'en velocity definitions (see
below), one might expect that the FLRs likely occur at $R_s < r <
100 R_s$.

Approximating the plasma inflow velocity with the free fall
velocity $v_p\sim v_{\rm ff}(r)$ and $v_{\rm A}\sim
B(r)/\sqrt{4\pi \rho_{\rm ff}}$ where $\rho_{\rm
ff}=\dot{M}/(v_{\rm ff}~ 4\pi r^2)$ is the free fall mass density,
one can rewrite Eq. (\ref{om_A}) as:
 \st \bea \omega^{\pm}(r)&\simeq&  k_z
\lp\frac{}{}
(2GM/r)^{1/2} \pm (B(r)^2/4\pi \rho_{\rm ff}(r)~)^{1/2}\frac{}{}\rp,\no\\
&\simeq& k_z(v_p(R_s) x^{-1/2} \pm v_{\rm A}(R_s) x^{-9/4})\,,
\eea where $v_p(R_s)=(2GM/R_s)=1.6\times 10^{10}
(M/M_\odot)^{1/2}(R_s/10\,{\rm km})^{-1/2}$ cm s$^{-1}$ and
$v_{\rm A}(R_s)=B/\sqrt{4\pi\rho_{\rm ff}(R_s)}\simeq 3\times
10^{10} \mu_{26} \dot{M}_{17}^{-1/2} (M/M_\odot)^{1/4}
(R_s/10\,{\rm km})^{-9/4}$ cm s$^{-1}$ are inflow and Alfv\'en
velocities at the surface of the star\footnote{We note that to
calculate the Alfv\'en velocity at the surface of the star
properly, one needs to use the relativistic Alfv\'en velocity as
defined in Eq. (\ref{rev_alfv}), ie. \st \be v_{\rm A}=\frac{\mu
\dot{M}^{-1/2}(2GM)^{1/4}r^{-9/4}}{\sqrt{1+\mu^2
\dot{M}^{-1}(2GM)^{1/2}r^{-9/2}/c^2}}.\ee  Using the above
relation the Alfv\'en velocity at the surface of the star will be
$\sim 0.8 c$, $.997 c$, and $c$ for $\mu=10^{26},\,10^{27}$, and
$10^{28}$ G cm$^3$, respectively.  In further distances from the
star, however, the relativistic Alfv\'en velocity reduces to its
classical version.}. Here $\mu_{26}$ is the magnetic field dipole
moment at the surface of star in units of $10^{27}$ G cm$^{-3}$,
$\dot{M}_{17}$ is the mass of accretion rate in units of $10^{17}$
g s$^{-1}$, $x=r/R_s$, and $R_s$ is the radius of star. Figures
\ref{freq_vs_r} and \ref{freq_vs_alfv} show the variation of
purely Alfv\'en resonance frequencies ($v_p=0$) or FLRs, as a
function of distance from the star and magnitude of Alfv\'en
velocity. It is clear that the closer to the star and/or the
larger the Alfv\'en velocity the higher the frequency. Therefore,
for an Alfv\'en velocity say $v_{\rm A}\sim .1 c$ one can get a
frequency about $1000$ Hz at $r\simeq 2 R_s$ as we expected.
Furthermore, since the Alfv\'en velocity depends on the enthalpy
(or density) of the plasma which varies from time to time due to
several processes such as magneto-turbulence at boundaries during
the accretion, the resulting FLRs frequencies will also vary with
time.  In addition, the position that the resonance takes place,
is also subject to change in time due to the time varying
accretion rate.   Such time varying behavior causes time varying
frequencies or quasi-periodic frequencies as observed.

Setting $\omega^-$ as the observed peak separation frequency
$\Delta \nu\sim \omega^-$, and $\omega^+$ as the upper QPO
frequency $\nu_2\sim\omega^+$, we can compare our results with
observations.  Figures \ref{peak_freq_low} and \ref{peak_freq}
show the variation of $\Delta\nu$ vs $\nu_1$ and $\nu_2$,
respectively, along with the observed values. The curves are drawn
by assuming the ratio $v_{\rm A}/v_p\simeq .8$ ar $r=R_s$ where
from bottom to top correspond to upper frequency $\nu_2\sim 1170$
Hz, $1350$ Hz, and $1548$ Hz at $r=R_s$, respectively. As shown by
observations, the value of $\Delta\nu$ decreases whenever the
magnitude of $\nu_2$ decreases (sufficiently) and/or it increases.
Such behavior is expected in our model (see Fig. \ref{peak_freq}).

Furthermore, the lower QPO frequency can be obtained from
$\nu_1=\nu_2 - \Delta\nu\sim 2k_z v_{\rm A}$.  As a result,
$\nu_1$ to $\nu_2$ we obtain the frequency ratio \st
\be\label{om_ratio} \nu_1/\nu_2= 2v_{\rm A}/(v_p + v_{\rm A})=
2/(1+v_p/v_{\rm A}) \,. \ee It is clear that depending on the
relation between the plasma flow velocity $v_p$ and the Alfv\'en
velocity $v_{\rm A}$, one can obtain different frequency ratios.
For example, if the velocity inflow plasma is $2$ ($7/3\sim 2.3$)
times bigger than the Alfv\'en velocity, ie. $v_p=2v_{\rm A}$, one
can get $2:3$ ($3:5$) ratio. For a free falling plasma along the
lines of a dipolar magnetic field, $v_p^2(r)\propto r^{-1}$,
whereas $B^2(r)/4\pi\rho(r)\propto r^{-9/2}$, the condition
$v_p\sim 2v_{\rm A}$ ($v_p\sim 2.3v_{\rm A}$) is satisfied at
$r\sim 3 R_s\sim 1.5 r_{\rm A}$. The occurrence of FLRs is very
likely at this distance from the star.

Such pairs of high frequencies with $2:3$ frequencies ratio have
been observed in the X-ray flux of almost all neutron stars in
LMXBs\footnote{In Sco X-1 the correlation line between two
frequencies has a steeper slope than $2/3$.} and black hole
systems (the $3:5$ ratio has seen in one source), and it would
appear this feature is common to those systems. However, the
existence of such rational ratios is still a mystery. In black
hole systems, it has been suggested that such frequencies
correspond to a trapped g-mode or c-mode of disk oscillation in
the Kerr metric, see for example \cite{Kat01} and references
therein.  In neutron star systems they are explained as the
fundamental and the first harmonic of the non-axisymmetric ($m=1$)
g-mode \citep{Kat02,Kat03}. Further, \citet{Rez03a,Rez03b} studied small
perturbations of an accretion
torus orbiting close to the black hole and modelled the observed
high QPO frequencies with basic p-modes of relativistic tori.
They showed that these modes behave as sound waves trapped in
the torus with eigenfrequencies appearing in sequence 2:3:4:...
\cite{Abr03} also proposed that
the observed rational ratios of frequencies may be due to the
strong gravity of the compact object and a non-linear resonance
between radial and vertical oscillations in accretion disks.


\section{Discussion}\label{conc}

In the present work, we have studied the interaction of an
accretion disk with a neutron star magnetosphere in the LMXBs. The
recent extensive observations reveal the existence of
quasi-periodic oscillations in the X-ray fluxes of such stars.
These oscillations, with frequencies range from $10$ Hz to $1200$
Hz, have been the subject of several theoretical and observational
investigations. Based on theoretical models for the observed
aurora in the earth magnetosphere, we have introduced a generic
magnetospheric model for accretion disk-neutron star systems to
address the occurrence and the behavior of the observed QPOs in
those systems. In order to explain those QPOs consistently, we
consider the interaction of the accreting plasma with the neutron
star' magnetosphere.   Due to the strong gravity of the star, a
very steep and supersonic flow hits the magnetosphere boundary and
deforms its structure drastically. Such a plasma flow can readily
excite different MHD waves in the magnetosphere, such as the shear
Alfv\'en waves.

\begin{figure}[h]
\begin{center}
\includegraphics[angle=0,width=12cm]{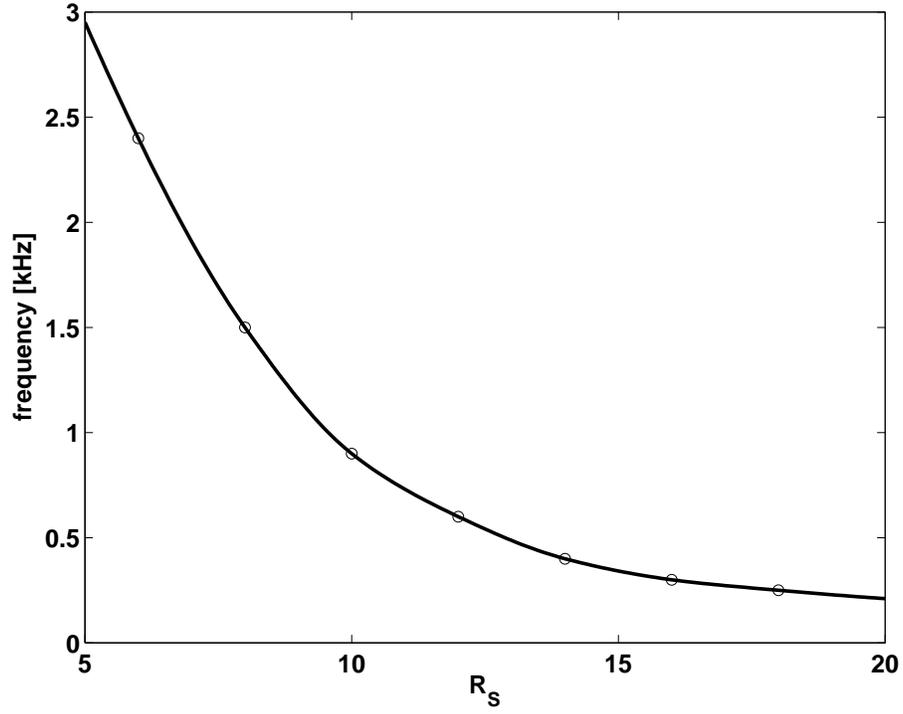}
\caption{Radial profile of the frequency of a FLR. The closer to
the star the higher the frequency. Here $v_{\rm A}=.5c$.}
\label{freq_vs_r}
\end{center}
\end{figure}

\begin{figure}[h]
\begin{center}
\includegraphics[angle=0,width=12cm]{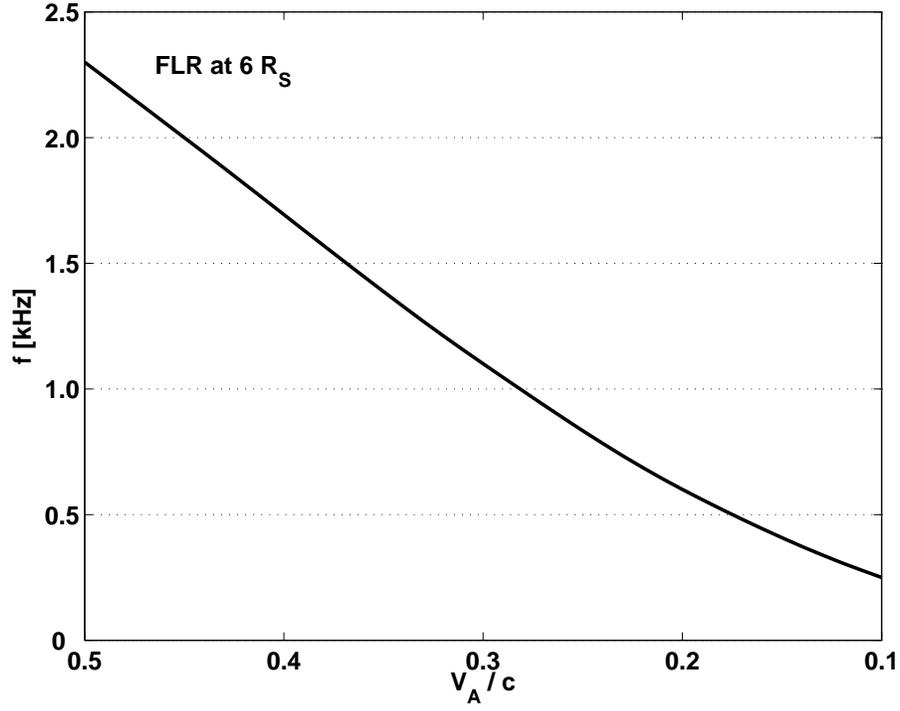}
\caption{Frequency of a FLR at $6R_s$ from the star as function of
Alfv\'en wave velocity. The larger the Alfv\'en velocity the
higher the frequency. } \label{freq_vs_alfv}
\end{center}
\end{figure}

\begin{figure}[h]
\begin{center}
\includegraphics[angle=0,width=12cm]{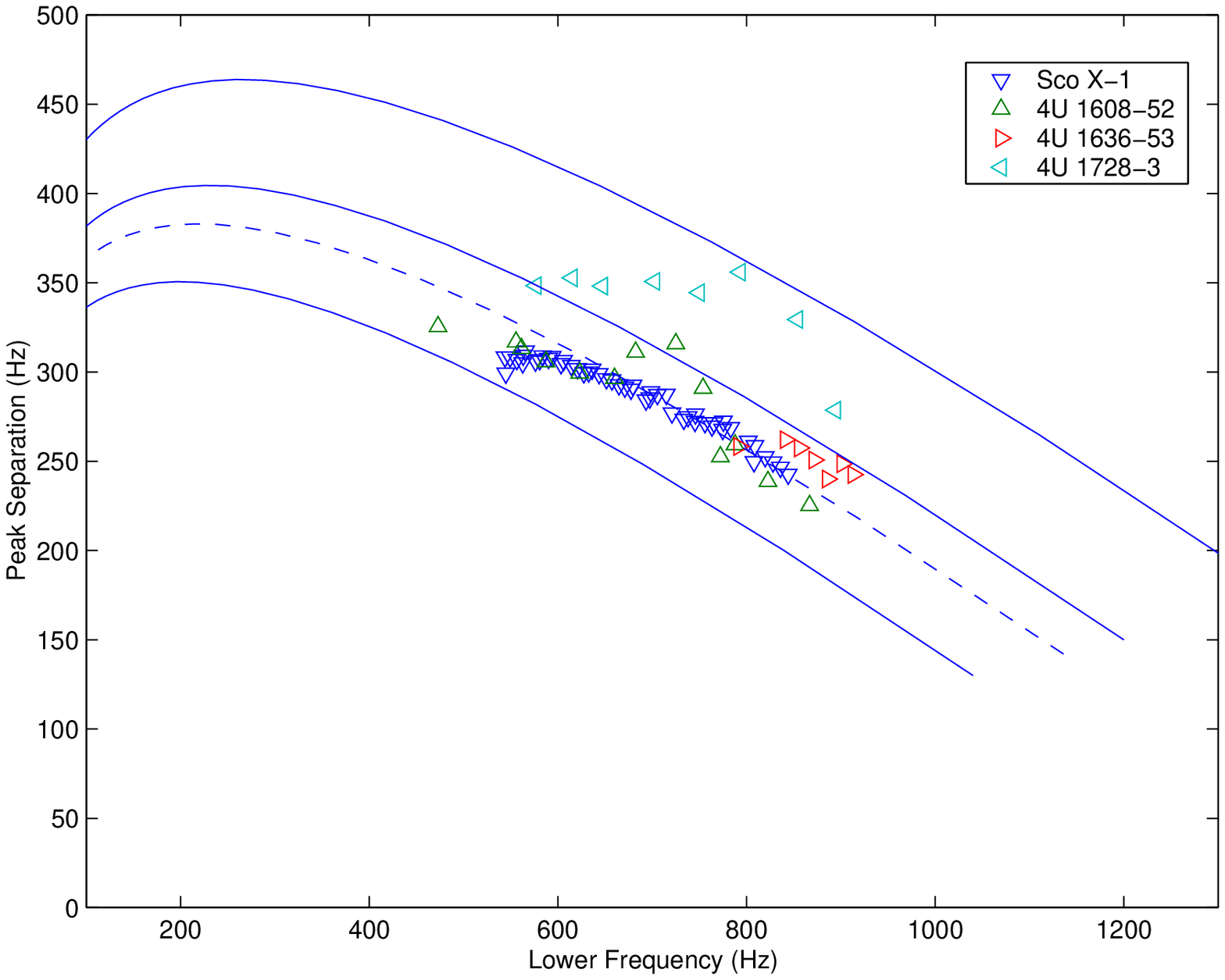}
\caption{Variation of peak separation frequency $\Delta\nu$ with
the upper frequency $\nu_2$.  The resulting curves are compared
with the observed values for different system. The solid curves
from bottom to top correspond to $\nu_2\sim 1170,\,1350,$ and
$1548$ Hz at $r=R_s$. The dashed curve corresponds to $\nu_2=1278$
Hz at $r=R_s$. } \label{peak_freq_low}
\end{center}
\end{figure}

\begin{figure}[h]
\begin{center}
\includegraphics[angle=0,width=12cm]{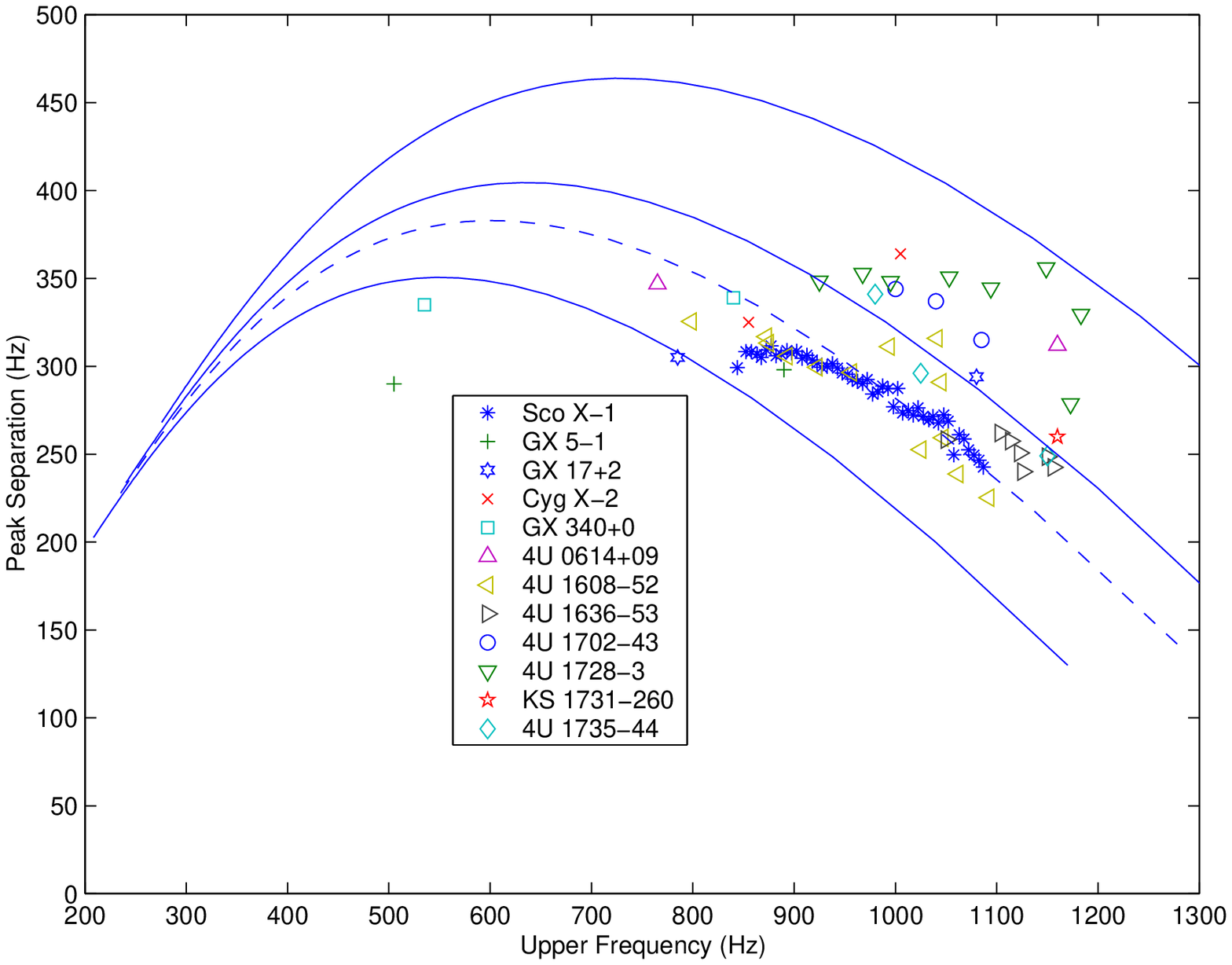}
\caption{Variation of peak separation frequency $\Delta\nu$ with
the upper frequency $\nu_2$.  The resulting curves are compared
with the observed values for different system. The solid curves
from bottom to top correspond to $\nu_2\sim 1170,\,1350,$ and
$1548$ Hz at $r=R_s$. The dashed curve corresponds to $\nu_2=1278$
Hz at $r=R_s$. } \label{peak_freq}
\end{center}
\end{figure}

In the Earth's magnetosphere, occurrence of aurora is a result of
resonant coupling between the shear Alfv\'en waves and
compressional waves (produced by the solar wind). These resonances
are known as FLRs that are reviewed in detail in section
\ref{sect:mag} of this paper. We argue that such resonant coupling
is likely to occur in neutron star magnetospheres during its
interaction with accreting plasma.  We formulated an improved FLR
by considering a plasma flow moving with velocity $v_p$ along the
magnetic field lines. Such flow is likely to occur in neutron star
magnetosphere \citep{GLP77}. For a simple geometry, the
rectilinear magnetic field, and in the presence of a plasma flow
we found: (a) two resonant hydrodynamical-MHD modes with
frequencies $\omega^{\pm} = k_z(v_p \pm v_{\rm A})$. (b) the
resulting frequencies for $k_z\sim \pi/R_s\sim 3\times 10^{-6}$
cm$^{-1}$ and typical flow and/or Alfv\'en velocity $\sim .1 c$
will be in kHz range. Our results match the kHz oscillations
observed in the X-ray fluxes in LMXBs. As shown in figures
\ref{freq_vs_r} and \ref{freq_vs_alfv}, the closer to the star
and/or the larger the Alfv\'en velocity the higher the frequency.
(c) the quasi-periodicity of the observed oscillations can be
understood by noting that due to several processes such as
magneto-turbulence at boundaries and the time varying accretion
rate, the FLR frequencies may vary with time. (d) a non-zero
plasma displacement along the magnetic field lines $\xi_{||}$.
Such a displacement that oscillates with resulting frequencies
$\omega^\pm$ modulates the flow of the plasma toward the surface
of neutron star.  As a result, the X-ray flux from the star will
show these frequencies as well. (e) setting $\Delta\nu=\omega^-$
and $\nu_2=\omega^+$, one can well explain the behavior of the
peak separation frequency $\Delta\nu$ relative to the upper QPO
frequency $\nu_2$.  As observed, the value of $\Delta\nu$
decreases as the magnitude of $\nu_2$ decreases and/or increases.
Figure \ref{peak_freq} clearly shows such behavior. (f) for
$v_p\sim 2 v_{\rm A}$ at $r\sim 3 R_s$, the frequency ratio
$\nu_1/\nu_2$ is comparable with the observed frequency ratio
$2:3$.

Interestingly, using the observed values of QPO frequencies, one
can determine the average mass density and the magnetic field
density of the star as
\st
\bea
\stq
&&k_z v_p(r)=(1/2)(\nu_2+\Delta\nu)=(\nu_2-\nu_1/2),\\
\stq
&&k_z v_{\rm A}(r)=(1/2)(\nu_2-\Delta\nu)=\nu_1/2\,.
\eea
Therefore, \st \bea \label{den}
\stq
&&M_s/R_s^3 \simeq (1/8\pi^2 G) x^3 (\nu_2+\Delta\nu)^2 ,\\
\stq &&\mu_s\simeq  (32\pi^4 G)^{-1/4} R_s^{5/2} \dot{M}^{1/2}
(M_s/R_s^3)^{-1/4} x^{13/4} \nu_1, \eea where $k_z\simeq \pi/r$,
$\mu_s=B_0R_s^3$,$x=r/R_s$, and $B_0$ is the magnetic field
strength at the surface of the star.   In table 1, we calculate
the average density $M_s/R_s^3$ and magnetic dipole moment $\mu_s$
of the star for a fixed value $x=10$. Note that larger frequencies
may occur at smaller distances, ie. $x<10$.  Our results are
comparable with realistic neutron star parameters.

Furthermore, our model is able to explain those low frequency
($\sim 10$ Hz) quasi-periodic oscillations observed in the rapid
burster such as MXB 1730-335 and GRO J1744-28 \citep{Mas00}.  Such
low frequencies can be extracted from the model by considering
smaller inflow/Alfv\'en velocity and/or further distances from the
star.

Nevertheless, in order to avoid a number of complexities in our
calculations, we used approximations such as slow rotation and
cold plasma.  These assumptions may put some restriction on the
validity of our model and results.  Future studies will be devoted
to overcoming those restrictions.

VR wishes to thank Mariano Mendez for kindly providing QPO data.
This research was supported by the National Sciences and
Engineering Research Council of Canada.

%
%
\def\aj{{AJ}}                   
\def\araa{{ARA\&A\ }}             
\def\apj{{ApJ\ }}                 
\def\apjl{{ApJ\ }}                
\def\apjs{{ApJS\ }}               
\def\apss{{Ap\&SS}}             
\def\aap{{A\&A\ }}                
\def\aapr{{A\&A~Rev.}}          
\def\aaps{{A\&AS}}              
\def\azh{{AZh}}                 
\def\baas{{BAAS}}               
\def\jrasc{{JRASC}}             
\def\memras{{MmRAS}}            
\def\mnras{{MNRAS\ }}             
\def\pra{{Phys.~Rev.~A}}        
\def\prb{{Phys.~Rev.~B}}        
\def\prc{{Phys.~Rev.~C\ }}        
\def\prd{{Phys.~Rev.~D\ }}        
\def\pre{{Phys.~Rev.~E}}        
\def\prl{{Phys.~Rev.~Lett.\ }}    
\def\pasp{{PASP}}               
\def\pasj{{PASJ\ }}               
\def\qjras{{QJRAS}}             
\def\skytel{{S\&T}}             
\def\solphys{{Sol.~Phys.}}      
\def\sovast{{Soviet~Ast.\ }}      
\def\ssr{{Space~Sci.~Rev.\ }}     
\def\zap{{ZAp}}                 
\def\nat{{Nature\ }}              
\def\iaucirc{{IAU~Circ. No.}}       
\def\aplett{{Astrophys.~Lett.}} 
\def\apspr{{Astrophys.~Space~Phys.~Res.}}
\def\bain{{Bull.~Astron.~Inst.~Netherlands}}
\def\fcp{{Fund.~Cosmic~Phys.}}  
\def\gca{{Geochim.~Cosmochim.~Acta}}   
\def\grl{{Geophys.~Res.~Lett.}} 
\def\jcp{{J.~Chem.~Phys.}}      
\def\jgr{{J.~Geophys.~Res.}}    
\def\jqsrt{{J.~Quant.~Spec.~Radiat.~Transf.}}
\def\memsai{{Mem.~Soc.~Astron.~Italiana}}
\def\nphysa{{Nucl.~Phys.~A}}   
\def\nphysb{{Nucl.~Phys.~B\ }}   
\def\physrep{{Phys.~Rep.}}   
\def\physscr{{Phys.~Scr}}   
\def\planss{{Planet.~Space~Sci.}}   
\def\procspie{{Proc.~SPIE}}   

----------------------------------------------------------------------------------------

\newpage

%
%
\begin{table}[htbp]
\caption{Observed frequencies of kilohertz QPOs in Z and atoll
sources} \label{table1} \scriptsize
\begin{center}
\begin{tabular}{lccclcc}
\hline
         &       &$\nu_1$ &$\nu_2$& $\Delta \nu$ & $M_s/R_s^3$ & $\mu_s$\\
Source   & Type  &(Hz)    & (Hz) & (Hz)& ($10^{14}$ g cm$^{-3}$)& ($10^{26}$ G cm$^3$)\\
\hline
Sco\,X$-$1     & Z  & 565 & 870  & 307$\pm$5  &  2.6  & 3.8  \\
GX\,5$-$1      & Z  & 660 & 890  & 298$\pm$11 &  2.7 & 4.1\\
GX\,17+2       & Z  & 780 & 1080 & 294$\pm$8  &  3.5  & 4.8\\
Cyg\,X$-$2     & Z  & 660 & 1005 & 346$\pm$29 &  3.4  & 4.1 \\
GX\,340+0      & Z  & 565 & 840  & 339$\pm$8  &  2.6  & 3.8 \\
GX\,349+2      & Z  & 710 & 980  & 266$\pm$13 &  2.9   & 4.5 \\
4U\,0614+09    & atoll  & 825    & 1160  & 312$\pm$2 &  4.1 & 4.8\\
4U\,1608$-$52  & atoll  & 865    & 1090  & 225$\pm$12 & 3.2 & 1.7\\
4U\,1636$-$53  & atoll  & 950    & 1190  & 251$\pm$4  & 3.9 & 5.4\\
4U\,1702$-$43  & atoll  & 770    & 1085  & 315$\pm$11 & 3.7 & 4.8\\
4U\,1705$-$44  & atoll  & 775    & 1075  & 298$\pm$11 & 3.5 & 4.8\\
4U\,1728$-$3   & atoll  & 875    & 1160  & 349$\pm$2  & 4.3 & 5.1\\
KS\,1731$-$260 & atoll  & 900    & 1160  & 260$\pm$10 & 3.8 & 5.4\\
4U\,1735$-$44  & atoll  & 900    & 1150  & 249$\pm$15 & 3.7 & 5.4\\
4U\,1820$-$30  & atoll  & 795    & 1075  & 278$\pm$11 & 3.4 & 4.8\\
Aql\,X$-$1     & atoll  & 930    & 1040  & 241$\pm$9  & 3.1 & 5.7\\
4U\,1915$-$05  & atoll  & 655    & 1005  & 348$\pm$11 & 3.4 & 4.1\\
XTE\,J2123$-$058& atoll & 845   & 1100   & 255$\pm$14 & 3.4 & 5.1\\
\hline
\end{tabular}
\end{center}

\vbox{ Low ($\nu_1$) and high ($\nu_2$) QPO frequencies with
corresponding peak separation ($\Delta\nu$) observed in Z and
atoll sources. Only one series data is given as example, for
further detail see \cite{Van00}. Values of $M_s/R_s^3$ and $\mu_s$
are calculated from Eq. (\ref{den}) for a fixed value $x=10$.  We
note that larger frequencies may be occurred at smaller distances,
$x<10$. }
\end{table}

\clearpage

\newpage
\clearpage

\clearpage

\clearpage

\clearpage

\clearpage

\clearpage

\clearpage

\clearpage

\end{document}